\documentclass[aps,prd,preprint,showpacs,showkeys,superscriptaddress]{revtex4}

\usepackage{latexsym}
\usepackage[dvips]{graphicx}

\begin{document}


\title{Thermodynamics of warped AdS$_3$ black hole in the brick wall method}

\author{Wontae Kim}
\email[]{wtkim@sogang.ac.kr}
\affiliation{Department of Physics, Sogang University, Seoul 121-742, Korea}
\affiliation{Center for Quantum Spacetime, Sogang University, Seoul 121-742, Korea}

\author{Edwin J. Son}
\email[]{eddy@sogang.ac.kr}
\affiliation{Department of Physics, Sogang University, Seoul 121-742, Korea}

\date{\today}

\begin{abstract}
The statistical entropy of a scalar field on the warped AdS$_3$ black
hole in the cosmological topologically massive gravity 
is calculated based on the brick-wall method, 
which is different from the
Wald's entropy formula giving the modified area law due to the
higher-derivative corrections
in that the entropy still satisfies the area law.  
It means that the entropy for scalar excitations on this background
is independent of higher-order derivative terms or
the conventional brick wall method has some limitations to take into
account the higher-derivative terms.
\end{abstract}



\maketitle

\section{Introduction}

The holographic principle in quantum gravity suggested by 't
Hooft~\cite{thooft:hol} and Susskind~\cite{susskind} shows that a
region with boundary of area $\mathcal{A}$ is fully described by no
more than $\mathcal{A}/4$ degrees of freedom; in other words, degrees
of freedom of a spatial region reside on its boundary (for a recent
review, see Ref.~\cite{bousso}). Supported by the generalized second
law of black hole thermodynamics~\cite{bekenstein,hawking}, the
entropy of a gravitational object is shown to be proportional to or
less than the area of its boundary (horizon for a black
hole)~\cite{bousso:CEB}. However, the beautiful area law,
$S=\mathcal{A}/4$, in black hole thermodynamics does not seem to hold when
higher curvature terms are taken into account, though the entropy is
still proportional to a local geometric density integrated over a
cross-section of the horizon~\cite{jm}. 
Using the Wald's formula~\cite{wald:formula}, one can see
that the higher curvature terms give some corrections to the area law:
\begin{equation}
S_c = \frac{\mathcal{A}}{4} + \Delta S_c, \label{S:cor}
\end{equation}
where $\Delta S_c$ comes from the higher curvature terms.

One of the interesting higher curvature terms in three dimensions 
is the gravitational Chern-Simons (GCS) term. 
The three-dimensional topologically massive
gravity (TMG)~\cite{djt}, which consists of the Einstein action and
the GCS action with coupling $1/\mu_G$,
\begin{equation}
\label{action} \\
S_{TMG} = \frac{1}{\kappa^2} \int_{\mathcal M} d^3x \sqrt{-g} \left(R+\frac2{\ell^2}\right) + \frac{\ell}{6\kappa^2\nu} \int_{\mathcal M} d^3x \sqrt{-g} \epsilon^{\lambda\mu\nu} \Gamma_{\lambda\sigma}^{\rho}\left[\partial_{\mu} \Gamma_{\nu\rho}^{\sigma} + \frac{2}{3}\Gamma_{\mu\tau}^{\sigma}\Gamma_{\nu\rho}^{\tau}\right],
\end{equation}
has been extensively studied
with a negative cosmological
constant $\Lambda=-\ell^{-2}$ to investigate 
physical modes and geometric solutions~\cite{lss,ss,cdww,gj,gkp,dt,hhkt,nutku,bc,alpss,cd},
where $\kappa^2 = 16\pi G_{3}$ is a three-dimensional Newton constant, 
$\nu$ is a dimensionless coupling related to the graviton mass
$\mu_G=3\nu/\ell$, and $\epsilon^{\lambda\mu\nu}$ is a
three-dimensional anti-symmetric tensor defined by
$\tilde{\epsilon}^{\lambda\mu\nu}/\sqrt{-g}$ with
$\tilde{\epsilon}^{012}=+1$.
It was conjectured that the cosmological TMG becomes chiral and the massive graviton disappears at the chiral point ($\mu_G\ell=1$)~\cite{lss,ss}, while it was argued that the massive graviton modes cannot be gauged away at the chiral point~\cite{cdww} and that there exists the ``logarithmic primary'' which prevents the theory being chiral within consistent boundary conditions~\cite{gj,gkp}.
However, there is an agreement that the entropy of the Ba\~nados-Teitelboim-Zanelli (BTZ) black hole~\cite{btz} as a trivial solution to Eq.~(\ref{action}) obeys the Cardy formula~\cite{cardy} based on the AdS/CFT correspondence and the contribution from the GCS term does not vanish but depends on the inner horizon~\cite{ss:ent,kl}.

So far, we have mentioned the BTZ black hole in the cosmological TMG.
Remarkably, it was satisfied with both the Einstein
tensor (with the cosmological constant) and the Cotton tensor derived from
the GCS action, independently. Next, one may consider another kind of black hole
solution satisfying the equation of motion totally. For
instance, the warped anti-de Sitter (warped AdS$_3$)~\cite{ikop} 
can be considered as a candidate of new vacuum, while the
well-known AdS$_3$ is a vacuum of the BTZ black hole.    
Once the warped vacuum is defined, then the warped AdS$_3$ black hole
can be obtained by coordinate transformations.    
In fact, the warped AdS$_3$ geometry can be viewed 
as a fibration of the real line with a constant warp factor over
AdS$_2$, 
which reduces the $SL(2,R)_{L}\times SL(2,R)_{R}$ isometry group 
to $SL(2,R)\times U(1)$.
One of the solutions, which is free from
naked closed timelike curves (CTCs), is
the spacelike stretched black hole which has been  
studied in Refs.~\cite{bc,alpss}.
Actually, the other warped solutions are unphysical so that one can consider
the spacelike stretched warped geometry of $\nu^2>1$ without CTCs for some relevant
calculations~\cite{ok}. 
Especially for $\nu=1$, 
it is just the BTZ solution where the asymptotic geometry is AdS$_3$.
The apparent metric will not be the same with the standard form which
will be discussed as a special case in the end of section III.  

Recently, the entropy of the warped AdS$_3$
black hole has been studied and showed that 
the entropy~(\ref{S:cor}) receives some corrections~\cite{alpss}. 
Intriguingly, 
it does not satisfy the area law; specifically, 
the entropy correction is given by 
$\Delta S_c=-(\pi/24\nu G_3)
\left[ 3(\nu^2-1)r_+ + (\nu^2+3)r_- - 2\nu\sqrt{r_+r_-(\nu^2+3)} \right]$,
where $r_\pm$ are radial coordinates of inner and outer horizons. 
In addition, it has been conjectured 
that the entropy satisfies the formula 
for the entropy of a two-dimensional CFT at temperatures $T_L$ and $T_R$,
$S_c = (\pi^2\ell/3) \left( c_L T_L + c_R T_R \right)$,
defining appropriate left and right moving central charges $c_L$ and $c_R$. The right moving central charge $c_R$ has been recently calculated in Ref.~\cite{cd}; however, the authors obtained the conjectured one with the opposite sign, where the negative central charge is supported by the fact that the corresponding Virasoro generator $L_0$ is bounded from below.

Actually, the above entropy correction in Ref.~\cite{alpss} is
due to the Wald entropy formula which is essentially based on the thermodynamic first
law~\cite{wald:formula}. The first law is actually 
related to the conserved charges through the Noether theorem. 
So, the action may directly affect the conserved charges, especially
entropy. On the other hand, it has been believed 
that the brick wall calculation~\cite{thooft} 
is a useful method to get the statistical black hole entropy by
counting the possible modes of quantum gas in the
vicinity of the horizon so that 
the entropy of various black holes can be expressed by the area law~\cite{thooft,mcguigan,lkk}.  
In particular, the entropy of the BTZ black hole
has been shown to satisfy the area law~\cite{lkk}. 
So, we can expect the Bekenstein-Hawking's area law as long as we use
the brick wall method even in the presence of the higher-derivative
term, since the brick wall method depends only on the metric rather
than the action.    

So, in this paper, we would like to consider the spacelike warped AdS$_3$ black
hole and study the entropy by using 
the brick wall method whether it really gives the area law or not.   
To this end, in Sec.~\ref{sec:free}, 
we first obtain the free energy by carefully considering the superradiance
in the rotating black hole.  
Then, thermodynamic quantities for the spacelike stretched warped
AdS$_3$ solution are
calculated in Sec.~\ref{sec:thermo}. 
In this brick wall formulation, we will obtain the Bekenstein-Hawking's area law
without higher-derivative corrections to the area law, 
and find the black hole system is thermodynamically stable since the
corresponding heat capacity is positive. It means that 
the brick wall entropy is independent 
of higher-order derivative terms. 
Finally, in
Sec.~\ref{sec:discuss}, 
discussions and comments will be given.

\section{Free energy for a warped AdS$_3$ black hole}
\label{sec:free}

\begin{figure}[pbt]
\begin{center}
  \includegraphics[width=0.5\textwidth]{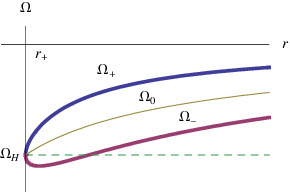}
  \caption{\label{fig:O}The upper and lower thick solid curves
    represent $\Omega_+$ and $\Omega_-$ with respect to
    $r\in(r_+,\infty)$, respectively. The middle solid curve is
    $\Omega_0$, and the horizontal dashed line is $\Omega_H$. 
It can be shown that all curves are negative, which means that all observers on timelike trajectories have negative angular velocities, since $\Omega_-<\Omega_{ob}<\Omega_+$.}
\end{center}
\end{figure}

Varying the cosmological TMG action~(\ref{action}), the bulk equation of motion is obtained as
\begin{equation}
G_{\mu\nu} - \frac1{\ell^2} g_{\mu\nu} + \frac\ell{3\nu} C_{\mu\nu} = 0,
\end{equation}
where $G_{\mu\nu}$ is the Einstein tensor and $C_{\mu\nu}$ is the Cotton tensor defined by
\begin{equation}
C_{\mu\nu} = \epsilon_\mu^{~\alpha\beta} \nabla_\alpha \left( R_{\beta\nu} - \frac14 g_{\beta\nu} R \right).
\end{equation}
A nontrivial solution in which the Cotton tensor does not vanish
is the spacelike stretched warped AdS$_3$ black hole given by~\cite{bc,alpss}
\begin{equation}
\label{met}
ds^2 = - N^2(r) dt^2 + R^2(r) (d\theta - \Omega_0(r) dt)^2 + \frac{\ell^2 dr^2}{4N^2(r)R^2(r)},
\end{equation}
where the functions in the metric are defined as
\begin{eqnarray}
N^2(r) &=& \frac{(\nu^2+3)(r-r_+)(r-r_-)}{4R^2(r)}, \\
R^2(r) &=& \frac{r}{4} \left[ 3(\nu^2-1)r + (\nu^2+3)(r_++r_-)-4\nu\sqrt{r_+r_-(\nu^2+3)} \right], \\
\Omega_0(r) &=& -\frac{2\nu r-\sqrt{r_+r_-(\nu^2+3)}}{2R^2(r)}.
\end{eqnarray}
Here, we choose $x^\mu=(x^0,x^1,x^2)=(t,\theta,r)$, and the coordinates $t$ and $r$ are dimensionful, while in Ref.~\cite{alpss}, they are dimensionless, i.e.\ 
$t\rightarrow\ell t$ and $r\rightarrow\ell r$.
The radial coordinates $r_\pm$ represent the inner and outer horizons, 
and $\Omega_0$ is the angular velocity of
zero-angular-momentum-observer (ZAMO); 
for instance, $\Omega_0 \to \Omega_H = -2\left[2\nu
  r_+-\sqrt{r_+r_-(\nu^2+3)}\right]^{-1}$ 
for $r\to r_+$.
In connection with the angular velocity, 
it is interesting to note that 
there is no stationary observer in our spacetime 
because $\Omega_+\Omega_-$ does not vanish except the infinity,
$\Omega_+\Omega_-=1/R^2>0$, where $\Omega_\pm \equiv \Omega_0 \pm N/R$
represent limits of angular velocity, i.e. for an observer,
$\Omega_-<\Omega_{ob}<\Omega_+<0$ (see Fig.~\ref{fig:O}). 
The given metric~(\ref{met}), however, describes the spacetime seen by a rest observer at infinity (ROI) because both $\Omega_\pm$ vanish at the infinity ($r\to\infty$).

Now, let us solve a scalar field equation of motion in this background
metric. 
Assuming $\Phi = e^{-i\omega t}e^{im\theta}\phi(r)$, 
the radial part of the Klein-Gordon equation
$[\Box-\mu^2]\Phi(t,\theta,r)=0$ yields
\begin{equation}
\left[ N^{-2}R^{-2} \frac{d}{dr} N^2R^2 \frac{d}{dr} + k^2(r;\omega,m) \right] \phi = 0, \label{KG:red}
\end{equation}
where
\begin{equation}
k^2(r;\omega,m) = \frac{\ell^2}{4} N^{-4}R^{-2} \left[ (\omega-\Omega_+m)(\omega-\Omega_-m) - \mu^2N^2 \right]. \label{k}
\end{equation}
Hereafter, we will consider only massless scalar field ($\mu=0$) for the sake of convenience.
Note that the wave number $k$ 
is dominant near horizon because 
it diverges when $r$ goes to $r_+$. 
It is thus sufficient to consider a thin-layer in the vicinity of the
horizon in order to obtain thermodynamic quantities~\cite{xz}.
In the WKB approximation, then, 
according to the semiclassical quantization rule with 
the periodic boundary condition $\phi(r+h)=\phi(r+h+\delta)$,
the total number of radial modes for a given energy $\omega$ is given by
\begin{equation}
\mathcal{N} = \sum_m n(\omega,m) = \frac1\pi \int dm \int_{r_++h}^{r_++h+\delta} dr\, k(r;\omega,m), \label{n}
\end{equation}
where the integration goes over those values for which $k^2\ge0$.
The cutoff parameters $h$ and $\delta$ 
are assumed to be very small positive quantity
compared to the horizons $r_\pm$.

It is worth noting 
that the ingoing modes seen by a ZAMO near the horizon given by
$\Phi_\mathrm{in}^\mathrm{ZAMO} \sim \exp \left[ i\tilde\omega \tilde{t} - i
  \tilde{m} \tilde\theta + i \int d\tilde{r}\, \tilde{k}(\tilde{r}) \right]$ are separated 
into the ingoing and outgoing modes seen by a ROI, $\Phi_\mathrm{in,out}^\mathrm{ROI} \sim \exp \left[ \pm i(\omega t - m\theta)+i \int dr\, k(r) \right]$,
where $\tilde{t} = t$, $\tilde{r} = r$, $\tilde\theta = \theta - \Omega_H t$, $\tilde
\omega = |\omega - \Omega_H m| >0 $, and $\tilde{m} = {\rm sgn}(\omega -
\Omega_H m)\, m$. Here, ${\rm sgn}(x)$ is 1 for $x > 0$ and $-1$ for $x <
0$. The superradiant (SR) modes are the modes with
$\tilde\omega=-(\omega - \Omega_H m)>0$, 
which are ingoing for the ZAMO near the horizon 
but outgoing for the ROI, $e^{i(\tilde\omega\tilde{t}-\tilde{m}\tilde\theta)} = e^{-i(\omega t-m\theta)}$, while the nonsuperradiant (NS) modes are the modes with
$\tilde\omega=(\omega - \Omega_H m)>0$, which are ingoing both for the
ZAMO near the horizon and the ROI,
$e^{i(\tilde\omega\tilde{t}-\tilde{m}\tilde\theta)} = e^{i(\omega
  t-m\theta)}$. 
The similar argument can be made for the outgoing modes 
for a ZAMO near the horizon, $\Phi_\mathrm{out}^\mathrm{ZAMO} \sim
\exp \left[ -i\tilde\omega \tilde{t} + i \tilde{m} \tilde\theta + i \int d\tilde{r}\,
  \tilde{k}(\tilde{r}) \right]$. 

Let us assume that this system is in thermal
equilibrium at a temperature $T=\beta^{-1}$ and take into account the superradiance as 
in Ref.~\cite{lkk}. Then, 
the free energy is given by
\begin{eqnarray}
F &=& \beta^{-1} \int d\mathcal{N} \ln (1-e^{-\beta(\omega-\Omega_Hm)}) = F_{NS}^{(m>0)} + F_{NS}^{(m<0)} + F_{SR}, \\
F_{NS}^{(m>0)} &=& -\frac{\ell}{2\pi} \int_{r_++h}^{r_++h+\delta} dr\, N^{-2}R^{-1} \int_0^\infty dm \int_0^\infty d\omega \frac{\sqrt{(\omega-\Omega_+m)(\omega-\Omega_-m)}}{e^{\beta(\omega-\Omega_Hm)}-1} \nonumber \\
  & & -\frac\ell{2\pi\beta} \int_{r_++h}^{r_++h+\delta} dr\, N^{-2}R^{-2} \int_0^\infty dm\, m \ln (1-e^{\beta\Omega_Hm}), \label{NS:m>0} \\
F_{NS}^{(m<0)} &=& -\frac{\ell}{2\pi} \int_{r_++h}^{r_++h+\delta} dr\, N^{-2}R^{-1} \int_{-\infty}^0 dm \int_{\Omega_-m}^\infty d\omega \frac{\sqrt{(\omega-\Omega_+m)(\omega-\Omega_-m)}}{e^{\beta(\omega-\Omega_Hm)}-1}, \label{NS:m<0} \\
F_{SR} &=& -\frac{\ell}{2\pi} \int_{r_++h}^{r_++h+\delta} dr\, N^{-2}R^{-1} \int_{-\infty}^0 dm \int_0^{\Omega_+m} d\omega \frac{\sqrt{(\omega-\Omega_+m)(\omega-\Omega_-m)}}{e^{-\beta(\omega-\Omega_Hm)}-1} \nonumber \\
  & & +\frac\ell{2\pi\beta} \int_{r_++h}^{r_++h+\delta} dr\, N^{-2}R^{-2} \int_{-\infty}^0 dm\, |m| \ln (1-e^{-\beta\Omega_Hm}). \label{SR}
\end{eqnarray}
After some tedious calculations, the above free energies can be reduced to
\begin{eqnarray}
F_{NS}^{(m>0)} &\simeq& \frac{\ell\zeta(3)(2\nu r_+-\sqrt{r_+r_-(\nu^2+3)})^3}{2\pi\beta^3(\nu^2+3)^{3/2}(r_+-r_-)^{3/2}} \left[ \frac\pi{\sqrt{h+\delta}}-\frac\pi{\sqrt{h}} \right] \nonumber \\
  & & -\frac{\ell\zeta(3)(2\nu r_+-\sqrt{r_+r_-(\nu^2+3)})^2}{2\pi\beta^3(\nu^2+3)(r_+-r_-)} \ln \frac{h+\delta}{h}, \\
F_{NS}^{(m<0)} &\simeq& -\frac{\ell\zeta(3)(2\nu r_+-\sqrt{r_+r_-(\nu^2+3)})^2}{\pi\beta^3(\nu^2+3)^2(r_+-r_-)^2} \ln \frac{h+\delta}{h} \nonumber \\
  & & \quad \times \left[ 3(\nu^2-1)r_+ + (\nu^2+3)r_- - 2\nu\sqrt{r_+r_-(\nu^2+3)} \right], \\
F_{SR} &\simeq& -\frac{\ell\zeta(3)(2\nu r_+-\sqrt{r_+r_-(\nu^2+3)})^3}{2\pi\beta^3(\nu^2+3)^{3/2}(r_+-r_-)^{3/2}} \left[ -\frac\pi{\sqrt{h+\delta}} + \frac\pi{\sqrt{h}} \right] \nonumber \\
  & & -\frac{\ell\zeta(3)(2\nu r_+-\sqrt{r_+r_-(\nu^2+3)})^2}{2\pi\beta^3(\nu^2+3)^2(r_+-r_-)^2} \ln \frac{h+\delta}{h} \nonumber \\
  & & \quad \times \left[ (\nu^2+3)(r_+-r_-) - 4\nu \left(2\nu r_+-\sqrt{r_+r_-(\nu^2+3)}\right) \right]
\end{eqnarray}
up to zeroth order in $h/r_+$ and $\delta/r_+$. 
It is easy to check that all the logarithmic terms are remarkably
canceled out, 
so that the total free energy is finally obtained as
\begin{eqnarray}
F &=& \frac{\ell\zeta(3)(2\nu r_+-\sqrt{r_+r_-(\nu^2+3)})^3}{\beta^3(\nu^2+3)^{3/2}(r_+-r_-)^{3/2}} \left[ \frac1{\sqrt{h+\delta}}-\frac1{\sqrt{h}} \right] \nonumber \\
  &\simeq& -\frac{2\ell\zeta(3)(2\nu r_+-\sqrt{r_+r_-(\nu^2+3)})^3}{\beta^3(\nu^2+3)^2(r_+-r_-)^2} \left(\frac\ell{\bar{h}} \right) \label{F}
\end{eqnarray}
in the leading order. Here, we have assumed $\bar{h}\ll\bar\delta$, where the proper lengths of cutoffs $\bar{h}$ and $\bar\delta$ are given by
\begin{eqnarray}
\bar{h} &\equiv& \int_{r_+}^{r_++h} dr\, \sqrt{g_{rr}} \simeq \frac{2\ell\sqrt{h}}{\sqrt{\nu^2+3}\sqrt{r_+-r_-}}, \\
\bar\delta &\equiv& \int_{r_++h}^{r_++h+\delta} dr\, \sqrt{g_{rr}} \simeq \frac{2\ell\left[\sqrt{h+\delta}-\sqrt{h}\right]}{\sqrt{\nu^2+3}\sqrt{r_+-r_-}},
\end{eqnarray}
respectively.
Note that the total free energy~(\ref{F}) blows up in the extremal
limit $(r_+-r_-)\to0$, which is not a defect, because we used the
approximation $h,\delta\ll(r_+-r_-)$ in the calculation so that
$(r_+-r_-)$ cannot vanish.

\section{Thermodynamic quantities}
\label{sec:thermo}
Let us calculate thermodynamic quantities by using the explicit form of the free energy.
First of all, the entropy can be obtained through 
the thermodynamic relation, $S=\beta^2(\partial F/\partial\beta)_\Omega|_{\beta=\beta_H} = -3\beta F|_{\beta=\beta_H}$, as
\begin{equation}
S = \frac{3\zeta(3)\mathcal{C}}{8\pi^3\ell} \left(\frac\ell{\bar{h}} \right), \label{ent}
\end{equation}
where the Hawking temperature $\beta_H^{-1}=\kappa_H/2\pi$ and the circumference of the horizon $\mathcal{C}$ are calculated as
\begin{eqnarray}
\beta_H^{-1} &=& \frac{(\nu^2+3)(r_+-r_-)}{4\pi\ell\left(2\nu r_+ - \sqrt{r_+r_-(\nu^2+3)}\right)}, \label{TH} \\
\mathcal{C} &\equiv& \oint_{r=r_+} d\theta \sqrt{g_{\theta\theta}} = \pi \left( 2\nu r_+ - \sqrt{r_+r_-(\nu^2+3)} \right),
\end{eqnarray}
and the surface gravity is given by $\kappa_H^2 = \left. -\frac12 \nabla^\mu\chi^\nu \nabla_\mu\chi_\nu \right|_{r=r_+}$ with the Killing vector $\chi^\mu = (\partial_t + \Omega_H\partial_\theta)^\mu$~\cite{wald:gr}.
Then, the entropy~(\ref{ent}) is exactly given by the one quarter area law,
\begin{equation}
S = \frac{2\mathcal{C}}{\ell_P}
  = \frac{2\pi\left(2\nu r_+-\sqrt{r_+r_-(\nu^2+3)}\right)}{\ell_P}, \label{S}
\end{equation}
where $\ell_P$ is the Planck length 
and we set the universal cutoff $\bar{h} = 3\zeta(3)\ell_P/16\pi^3$~\cite{thooft}. 
It is interesting to note that the present cutoff is the same universal constant
with that of the BTZ black hole.
The difference between the present entropy based on the brick wall method
and the statistical entropy $\mathcal{S}$ in Ref.~\cite{alpss} 
is explicitly written as  
\begin{eqnarray}
\mathcal{S} &=& \frac1{4G_3} \left[ \frac23 \mathcal{C} + \frac{\pi(\nu^2+3)}{6\nu} (r_+-r_-) \right] \nonumber \\
  &=& S - \frac1{24\nu G_3} \left[ 2\nu\mathcal{C} - \pi(\nu^2+3)(r_+-r_-) \right],
\end{eqnarray}
where we set $\ell_P = 8G_3$ to write a familiar form
of the area law $\mathcal{C}/4G_3$.
For convenience, let us define the entropy difference deviated from
the area law as 
\begin{equation}
\Delta S=\mathcal{S}-S=-\frac{\pi}{24\nu G_3} \left[ 3(\nu^2-1)r_+ + (\nu^2+3)r_- - 2\nu\sqrt{r_+r_-(\nu^2+3)} \right],
\end{equation}
which vanishes especially 
for $\sqrt{r_-/r_+}=(\nu\pm\sqrt{3-2\nu^2})/\sqrt{\nu^2+3}$.
The two different approaches yields the different result:
the brick wall method counts the number of modes 
of scalar field in the vicinity of the horizon, 
whereas the Wald's formula defines the entropy 
as a local conserved quantity by the Noether theorem.

Next, the angular momentum of the quantum gas $J_\mathrm{matter} = \left. -\left( \partial F/\partial \Omega \right)_\beta \right|_{\beta=\beta_H,\Omega=\Omega_H}$ is obtained as
\begin{eqnarray}
J_\mathrm{matter} &=& -\frac{1}{32\nu\ell\ell_P} \left[ \left(6(\nu^2-1)r_+ + (\nu^2+3)(r_++r_-) - 4\nu\sqrt{r_+r_-(\nu^2+3)}\right)^2 \right. \nonumber \\
  & & \qquad \qquad \left. - (\nu^2+3)^2(r_+-r_-)^2\right], \label{J}
\end{eqnarray}
in the leading order.
Then, the internal energy $U = F + \beta_H^{-1} S + \Omega_H J_\mathrm{matter}$ is simply given by,
\begin{equation}
U = \frac{1}{6\ell\ell_P} \left[ 6\nu \left(2\nu r_+-\sqrt{r_+r_-(\nu^2+3)}\right) - (\nu^2+3)(r_+-r_-) \right], \label{U}
\end{equation}
which is definitely positive.
In particular, 
the internal energy is written 
as $U\simeq\frac23\beta_H^{-1}S=(\nu^2+3)(r_+-r_-)/3\ell\ell_P$ 
for $\Delta S\simeq0$, because the angular momentum is actually
proportional to the entropy difference, $J_\mathrm{matter} = (3/4\pi\ell) \mathcal{C} \Delta S$, 
so that it is arbitrarily small in this case.
Finally, the heat capacity $C_J = (\partial U/\partial T)_J$ is calculated as
\begin{equation}
C_J = \frac{4\pi\left(2\nu r_+-\sqrt{r_+r_-(\nu^2+3)}\right)}{\ell_P},
\end{equation}
which is always positive and 
the spacelike warped AdS$_3$ black hole solution~(\ref{met}) is thermodynamically stable.

As a comment,  
in the limit of $\nu\to1$, the metric~(\ref{met}) corresponds to 
BTZ black hole as discussed in Ref.~\cite{alpss}.
However, the internal energy
$U=2(2\rho_+-\rho_-)/3\ell\ell_P=\left[\ell/(\rho_+-\rho_-)\right]
\left[U_\mathrm{BTZ}-2\rho_+\rho_-/\ell^2\ell_P\right]$ is not the
same with that of the BTZ black hole, 
while the entropy $S=4\pi \rho_+/\ell_P$ and the angular momentum
$J_\mathrm{matter}=2\rho_+\rho_-/\ell\ell_P$ are coincident with those
of the BTZ case, respectively, where $\rho_\pm\equiv
R(r_\pm)=\sqrt{r_\pm}\left(\sqrt{r_+}-\sqrt{r_-}\right)$ 
are the radii of the inner and outer horizon circles. 
Note that for a given internal energy 
$U = F + \beta_H^{-1} S + \Omega_H J_\mathrm{matter}$, 
the Hawking temperature $T_H =\left[\ell/(\rho_+-\rho_-)\right] T_H^\mathrm{BTZ}$ and the
angular velocity at the horizon
$\Omega_H=\left[\ell/(\rho_+-\rho_-)\right]
\left[\Omega_H^\mathrm{BTZ}-1/\ell\right]$ are different 
because the coordinate systems are different.
These differences can be explained as follows.
When $\nu=1$, there exists a coordinate transformation to the standard
form of the BTZ metric:
\begin{equation}
t = \frac{\rho_+-\rho_-}{\ell} \tau, \quad \theta = \varphi - \frac1\ell \tau, \quad r = \frac{\rho^2}{\rho_+-\rho_-}. \label{coord:trans}
\end{equation}
The transformed metric $ds^2 = -N_\mathrm{BTZ}^2 d\tau^2 +
N_\mathrm{BTZ}^{-2} d\rho^2 + \rho^2 \left( d\varphi -
  \Omega_\mathrm{BTZ} d\tau \right)^2$ is then exactly same as the BTZ
solution, where
$N_\mathrm{BTZ}^2=(\rho^2-\rho_+^2)(\rho^2-\rho_-^2)/\rho^2\ell^2$ and
$\Omega_\mathrm{BTZ}=\rho_+\rho_-/\rho^2\ell$ are the lapse function
and the angular velocity of ZAMO in the BTZ black hole, respectively.
It is then obvious that the
metric~(\ref{met}) in the limit of $\nu\to1$ describes the BTZ black
hole in the rotating frame and the different internal energy 
is related to the choice of the coordinate system.

\section{Discussions}
\label{sec:discuss}
We have shown that the entropy of the warped AdS$_3$ black hole 
can be calculated by using the 't Hooft's brick wall method so that it
gives
 the well-known area law of the black hole. 
It is interesting to note that the area law in the brick wall method
can be obtained generically as
long as one assumes,
$ds^2=-N^2dt^2+f^{-2}dr^2+R^2(d\theta-\Omega dt)^2$ where
$N^2=(r-r_+)\tilde{N}^2$, $f^2=(r-r_+)\tilde{f}^2$ with
$\tilde{N}^2(r_+)\ne0$ and $\tilde{f}^2(r_+)\ne0$. 
After some tedious calculations,
we can obtain the free energy, 
$F = -\left[4\zeta(3)R(r_+)/\beta^3\tilde{N}^2(r_+)\tilde{f}^2(r_+)\right] \bar{h}^{-1}$
in the leading order. Defining the Hawking temperature
$\beta_H^{-1}=\tilde{N}(r_+)\tilde{f}(r_+)/4\pi$, the entropy  
$S=-3\beta F|_{\beta=\beta_H}$ can be obtained as
$S = \left[6\zeta(3)\mathcal{C}/16\pi^3\right] \bar{h}^{-1}$,
where $\mathcal{C}=2\pi R(r_+)$ is the circumference of the horizon.
If $\bar{h}$ is independent of $r_\pm$, 
then the entropy is always proportional to the area of the horizon --- 
the horizon circumference in this (2+1)-dimensional case.
The higher derivative terms
in the action modify only the metric which is a solution of the modified
equation of motion. The scalar field feels the
geometry only through the deformed metric.
So, the expected entropy deformation does not appear at least in the
brick wall method.  

On the other hand, the entropy correction due to the higher derivative
terms such as GCS term in the present case,
do exist in the Wald formulation which is affected not only by the metric 
but also action itself. So, the Wald's formula seem to be more sensitive to the geometry
because the action can be used in the course of calculation.
Moreover, all conserved quantities are the Noether
charges. It is plausible that the
thermodynamic first law is automatically valid because it is just
Noether theorem in the Wald formulation. 
The Wald entropy is mathematically clear and
thermodynamically plausible so that it deserves to study statistically.
Unfortunately, the brick wall method to calculate the statistical entropy     
can not reproduce the Wald entropy.
Similar situation has been seen by considering only the GCS action with the BTZ black hole as a particular solution~\cite{ks}: The brick wall method gives the conventional area law with the outer horizon, while the Wald entropy is proportional to the area of the inner horizon.
It means that the brick wall entropy for scalar excitations is independent of
higher derivative term, otherwise the conventional brick wall method has some
limitations to take into account the higher-derivative contributions.
Further study is needed to clarify this issue.

Apart from the brick wall method, the other statistical calculation is
to use the Cardy formula in the dual CFT. The left-right central charges are
crucial to obtain the statistical entropy; however, the recent
calculation shows that the dual CFT is not unitary~\cite{cd}. 
So, it seems to be difficult to reproduce the Wald entropy statistically in the 
cosmological TMG.


\begin{acknowledgments}
This work was supported by the Korea Science and Engineering Foundation 
(KOSEF) grant funded by the Korea government(MOST) 
(R01-2007-000-20062-0).
\end{acknowledgments}


\end{document}